# A DFT computational design and exploration of novel direct band gap silver-thallium double perovskites

Syed Zuhair Abbas Shah[*,1,2,3], Shanawer Niaz[1,3], Tabassum Nasir[2], and James Sifuna[4,5]

[1]Department of Physics, Thal University Bhakkar, Bhakkar, 30000, Pakistan

[2]Institute of Physics, Gomal University, D.I. Khan, 29050, Pakistan

[3]Department of Physics, University of Sargodha, Sargodha, 40100, Pakistan

[4]Department of Natural Science, The Catholic University of Eastern Africa, 62157 - 00200, Nairobi, Kenya.

[5]Department of Technical and Applied Physics, The Technical University of Kenya, 52428-00200, Nairobi, Kenya.

*Email: zuhair.abbas@tu.edu.pk

*Tel: +92-300-7668414

## Abstract

Researchers have addressed the non-traditional power generation schemes as alternatives to the traditional fossil-fuel methods enormouslysincethescientific community has serious concerns aboutshortages of energy on our planet for future generations. In this scenario, innovative materials for photovoltaic and thermoelectric device applications are required by addressing current issues of instability and efficiency. Perovskites are very popular in this regard particularly having higher power conversion efficiency of 25.2% in thecase of solar cells. In thecurrent article, we investigated innovative small direct band gap double perovskites (elapsolite) $Cs_2AgTlX_6$ (X= Cl, Br) with a comprehensive discussion on structural, electronic, optical, and thermoelectric properties using afirst-principles approach. The compounds under investigation are found stable, efficient, and economical with alluring optical and thermoelectric properties. The higher absorption peaks in thevisible range, substantial optical conductivities ($\sim 10^{16}sec^{-1}$), and a lower percentage of reflection in thevisible range make these compounds fascinating for solar cell applications. Whereas large values of Seebeck coefficients, electrical conductivities, thefigure of merits (greater than unity), and smallvalues of thermal conductivities suggest the applications of these compounds in thermoelectric generators.

1. **Introduction:**

Keeping in view the energy shortages and ecological issues of our planet, the researchers are motivated to explore new energy harvesting mechanisms as alternatives to traditional energy sources. Solar and thermoelectric energy sources have gained much attention in thepast few years and extensive research has been conducted to explore efficiently and the state-of-the-art materials for such mechanisms so that their usage may be practiced as arenewable energy source on acommercial scale in thefuture. Both of the above-mentioned mechanisms (i.e.solar cells and thermoelectric generators) of alternative power production are cleaner, stable, noise-free, efficient, and cheaper compared with traditional fossil-fuel-based power generation schemes already implemented worldwide. The issues like efficiency and stability need to be addressed yet and hence thousands of researchers are continuously working in both experimental and theoretical domains to overcome the deficiencies.

Among various classes of materials, perovskites are one of the dominant classes whichhave been explored extensively for photovoltaic and thermoelectric applications, especially in the domain of perovskite solar cells. Perovskites are famous for their structural stability, reliability, tunable bandgap, high absorption coefficient, low effective mass, free availability, and cost-effectiveness, and interestingly, there is a probability of their synthesis in a desired structural form [1, 2]. Because of these promising characteristics, perovskites have numerous applications in photodetectors, lasers, light-emitting diodes (LEDs), X-rays, photo-catalysis, optoelectronic, and thermoelectric energy devices [3, 4]. Solar cell efficiencyis a burning issue among thescientific community and in the case of perovskite solar cells, it was reported at 3.8% in 2009 [5], ~15% till 2013 [6], and upto25.2% in 2019 [7-10].

Perovskites have multiple configurations with different general formulae like $ABX_3$, $A_2BX_6$, $A_3B_2X_9$,and $A_2BB'X_6$,etc. but in recent years double perovskites also famous with the name elapsolite [11] with general formulae $A_2BB'X_6$ (whereA is acation with oxidation +1, B is monovalent metal, B′ is a trivalent metal whereas X is a halogen) have been investigated by many researchers and very attractive results regarding optoelectronic and thermoelectric properties have been reported [12].Bismuth-based double perovskites $Cs_2AgBiCl_6$ and $Cs_2AgBiCl_6$ have been studied [13] and it was concluded that both of these materials have indirect band gaps with values of 2.77eV and 2.19eV respectively. This indirect nature of the band gap restricts their photovoltaic applications. To avoid this limitation, thallium (Tl) was substituted with bismuth (Bi) with its minimum concentration. This was found helpful in reducing the band gap and conversion of nature of the band gap i.e. transition from indirect to direct nature of the band gap [14].Antimony (Sb) based double perovskite has been synthesized and investigated experimentally [15] and it was observed that $Cs_2AgSbBr_6$ with a band gap of 1.64eV have reasonable optical properties. $Cs_2AgBiBr_6$ crystallizes in cubic form (space group #225) with a band gap of 1.60eV but the photovoltaic efficiency was sufficiently low hence it needs further improvements to be useful for optoelectronic applications on acommercial scale. Another halide elapsolite comprising silver and indium $Cs_2AgInCl_6$ was asemiconductor withadirect tunable band gap [16] having a value of 3.3eV and favorable for optoelectronic and thermoelectric properties.

From theliterature survey, it tends to be presumed that the vast majority of the halide perovskitestudied till now are wide band gap semiconductors ordinarily with band gaps greater than 2eV. It is observed that the band gap decreases if one replaces chloride with bromide and further decreases if one substitute iodide but overall band gaps remain higher than 2eV in most of the cases [17]. In light of the literature survey, it is established that the small direct band gap double perovskites can be attractive candidates for photovoltaic and thermoelectric device applications. Therefore, we decided to investigate thallium-based double perovskitestheoreticallywhich have already been synthesized[18].However, up to the best of our knowledge, the optical and thermoelectric properties of $Cs_2AgTlX_6$ (X= Cl, Br) have not been investigated thoroughly yet.In this study,we aimed to examine the structural, electronic, and opticalproperties of $Cs_2AgTlX_6$ (X= Cl, Br)using density functional theory (DFT) whereas the thermoelectric properties using solutions of DFT with Boltzmann transport theory. The article is organized as follows: The computational details are discussed in section 2, structural properties in section 3, electronic properties in section 4, optical properties in section 5, and thermoelectric properties in section 6. A brief summary of the results is described in section 7.

2. Computational details:

We conducted a theoretical study employingDFTto obtain Kohn-Sham states using pseudopotential plane-wave DFT code i.e.Quantum ESPRESSO [19]. The well-knowngeneralized gradient approximation (GGA) was employed. In particular, the exchange-correlation potential ofPerdew, Burke,and Ernzerhof (PBE) was utilized [20],and additionally Hubbard correction parameter U was also included to get rid of band gap underestimation by GGA and to improve the accuracy of other relevant properties since GGA is not adequate to include the exchange-correlation effects of highly localized *d* and *f* shell electrons [21]. The optimum value of theHubbard-U parameter was calculatedfrom density functional perturbation theory (DFPT) as implemented inQuantum ESPRESSO with the utilization of different q-point grids i.e. 4×4×4, 6×6×6, 8×8×8, and 10×10×10. The optimization of structures was performed via variable cell relaxation within the scheme of Broyden, Fletcher, Goldfarb, and Shanno(BFGS) in order to estimate the optimized lattice constants and atomic positions. The geometry optimizationwas performed until stress and forces were less than $10^{-4}$Ry/bohr$^3$ and $10^{-3}$ Ry/Bohr on all atomic sites. We used ak-points grid of 15×15×15 as per the Monkhorst-Pack scheme [22] along with norm-conserving pseudopotentials to achieve self-consistency and a further finer k-points grid of 25×25×25 was used where needed e.g. for density of states (DOS) and projected density of states (PDOS) calculations. The optimum values of 60 Ry and 600 Ry for kinetic energy cutoff of the wave function and particle density were utilized respectively. The value of convergence criteria for electrons was $1\times10^{-6}$ Ry and Gaussian smearing of 0.02 Ry was used throughout the calculations. Xcrysden [23] was used to obtain the coordinates of thehigh symmetry path of theirreducible Brillouin zone (Γ-X-W-K-L-W) for plotting band structures.

In thecase of optical properties,the dielectric function was calculated using Quantum ESPRESSO and other relevant optical parameters were calculated from this complex functionusing equations (3-9) presented in section 5.Moreover, We have determined the thermoelectric properties from Boltzmann transport theory with BoltzTrap code however the relaxation time τ was assumed constant (τ≈$10^{-14}$s) [24]. This is called constant relaxation time approximation (CRTA). As a matter of fact, CRTA has been extensivelypracticed in numerous theoretical investigations for thermoelectric materials and has anticipated enough good outcomes for temperature and carrier concentration for non-parabolic band structure thermoelectric materials [25-29].

## 3. Structural Properties:

Silver-thallium-based halide elapsolite $Cs_2AgTlX_6$ (X= Cl, Br) being investigated, in this study, have space group $Fm\bar{3}m$ (225) belonging to the face-centered cubic crystal structure (FCC) that is in accordance with the reported experimental data [18]. Both of the compounds contain ten atoms in the primitive unit cells whereas forty atoms in conventional cells. The occupied Wyckoff positions by elements Cs, Ag, Tl, and X (X=Cl & Br) are 8c, 4a, 4b, and 24e, respectively. The unit cell and crystal structure are shown in Fig. 1 (a) and (b).

The calculated values of lattice constants through variable cell relaxation in the current study using the DFT+U approach are very close to the experimental values with very little underestimation as expected. The detailed comparison is represented in Table 1.

**Table 1:** Lattice constants comparison with reported experimental data of elapsolite $Cs_2AgTlX_6$ (X= Cl, Br) and optimized Hubbard-U parameters from the first-principles approach.

| Compound | Approach | Equilibrium Lattice constants (Å) | Hubbard-U parameter (eV) | References |
|---|---|---|---|---|
| $Cs_2AgTlCl_6$ | GGA+U | 10.42 | 7.94 | 10.55(exp.) [18] |
| $Cs_2AgTlBr_6$ | GGA+U | 10.91 | 8.04 | 11.08 (exp.) [18] |

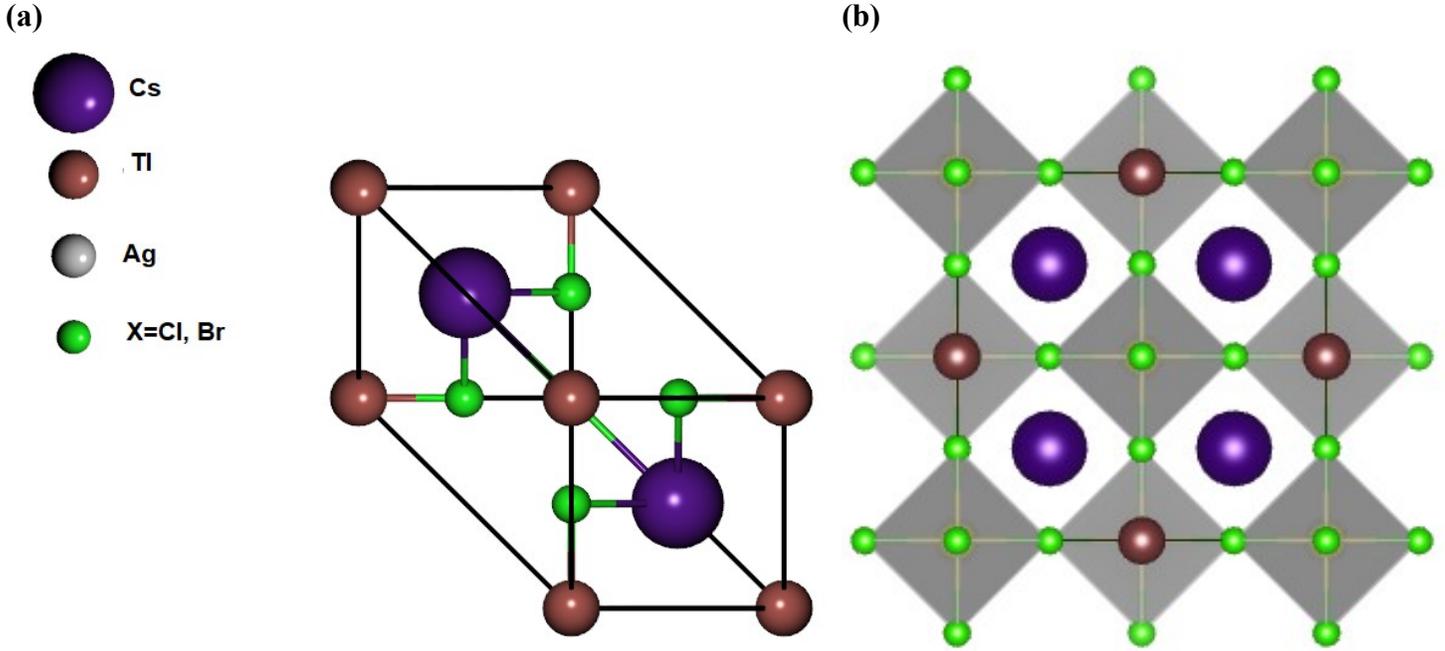

**Fig.1:** (a) Unit cell and (b) Crystal structure of elapsolite Cs$_2$AgTlX$_6$ (X= Cl, Br).

The assessment of structural stability of Cs$_2$AgTlX$_6$ (X= Cl, Br) was made by Goldschmidt's tolerance ($\tau_G$) and octahedral factor ($\mu$) which are very successful in predicting stable perovskite structures [30]. These factors were calculated using the following equations:

$$\tau_G = \frac{R_A + R_X}{\sqrt{2}(R_B + R_X)} \quad (1)$$

$$\mu = \frac{R_B}{R_X} \quad (2)$$

Considering the general formula of double perovskites A$_2$BB′X$_6$, $R_A$ be the ionic radius of cation A, $R_B$ be the average of two cations B and B′ radii whereas $R_X$ be the radius of the halogen ion. The calculated values of $\tau_G$ are 0.85 and 0.81 while that of $\mu$ are 0.69 and 0.63 for Cs$_2$AgTlCl$_6$ and Cs$_2$AgTlBr$_6$, respectively. The range of $\tau_G$ and $\mu$ proposed for a structure to be stable is 0.813 <$\tau_G$< 1.107 and 0.377 <$\mu$< 0.895 [31]. Therefore, it is clear that the calculated values predict the reasonable structural stability of the compounds under study.

4. **Electronic Properties:**

Electronic properties of materials play a key role in understanding carrier transport mechanisms thus enabling one to distinguish between metals, semiconductors, and insulators by analyzing the band gaps. These propertiesexhibit significant consequences on further properties such as optical and thermoelectric for theusage of material in various technological and commercial applications,for example, transistors, thermoelectric generators, and solar cells [32]. The electronic band structures of $Cs_2AgTlCl_6$ and $Cs_2AgTlBr_6$ have been calculated using the GGA+U method with high symmetry path Γ-X-W-K-L-W, shown in Fig.2. The exchange-correlation functional PBE underestimates the band gap [33] that's why we applied Hubbard-U correction for better estimation of band gaps. The band gap values along with acomparison with available reported experimental/ theoretical data are presented in Table 2.The compounds under study are inthe ideal range of band gaps i.e. 0.6-1.7 eV for solar cell applications [34]. Moreover, the band gap values of both compounds under investigation are direct (Γ-point) which is very fascinating for these materials to be regarded as attractive materials, particularly in solar cells and thermoelectric device applications [16].

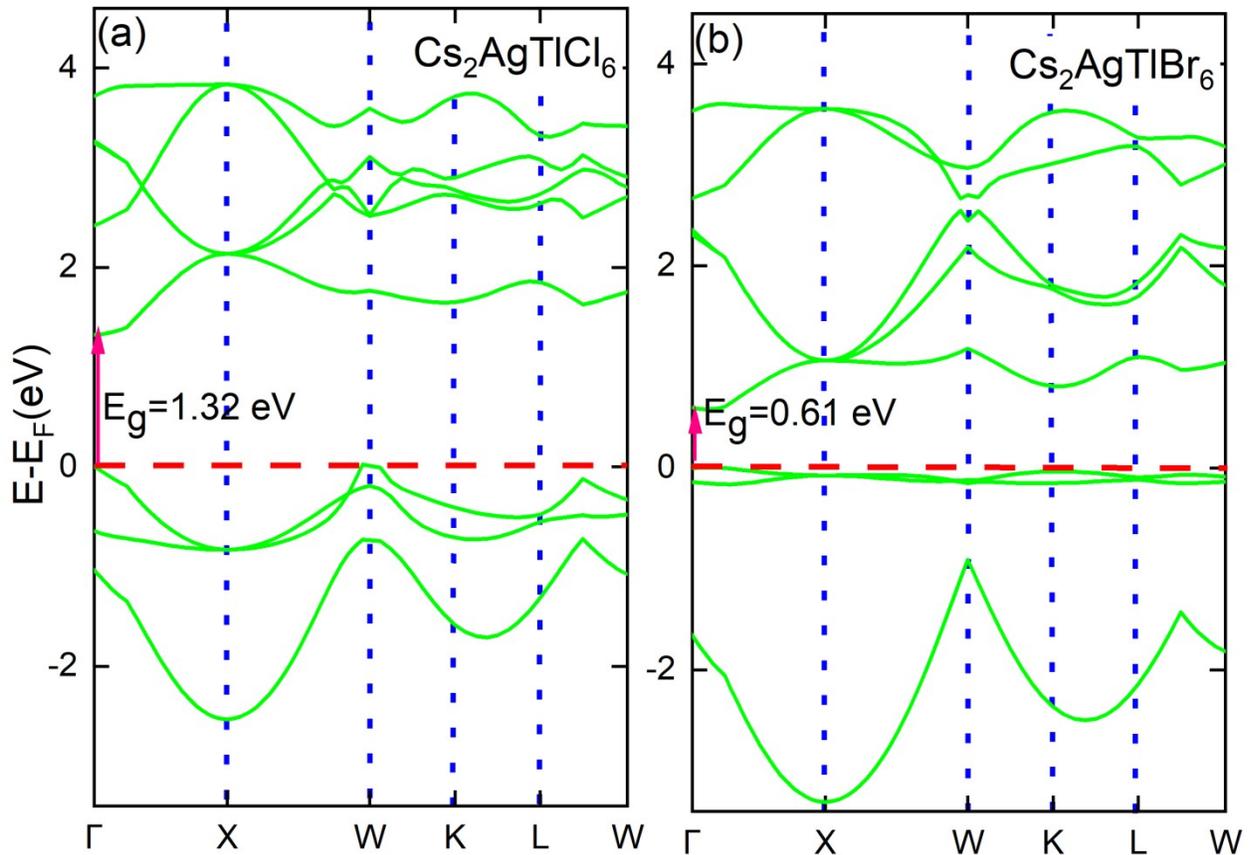

**Fig.2:** Calculated band-structures of elapsolite (a) $Cs_2AgTlCl_6$ and (b) $Cs_2AgTlBr_6$ by GGA+U method.

Moreover, DOS and PDOS were also calculated to describe, in detail, the elemental and orbital contribution in electronic structures. In general, DOS accounts for electronic states strength per unit volume per unit energy which is inversely related to the first derivative of E(k) with respect to wave vector k whilst effective mass is directly proportional to the curvature of E(k) with respect to k. PDOS depicts the elemental and orbital distribution of electrons within the solid. In the present case, DOS and PDOS of elapsolite $Cs_2AgTlCl_6$ and $Cs_2AgTlBr_6$ are represented in Fig.3 and Fig. 4 where total, element-wise, and orbital-wise contributions are presented. It is clear from Fig.3, Cs-s, Tl-d, and Cl-p orbitals actively contribute to the valence band residing on the left side of the Fermi level as the Fermi level is set to zero for convenience. On the other hand, Cs-p, Cl-p, and Ag-d orbitals contribute actively in the conduction band in the case of $Cs_2AgTlCl_6$ from Fig. 4. It is also clear that Cs-s, Ag-p, Tl-d, and Br-p make a major contribution in the valence band whereas Ag-p and Br-s in the conduction band in the case of $Cs_2AgTlBr_6$.

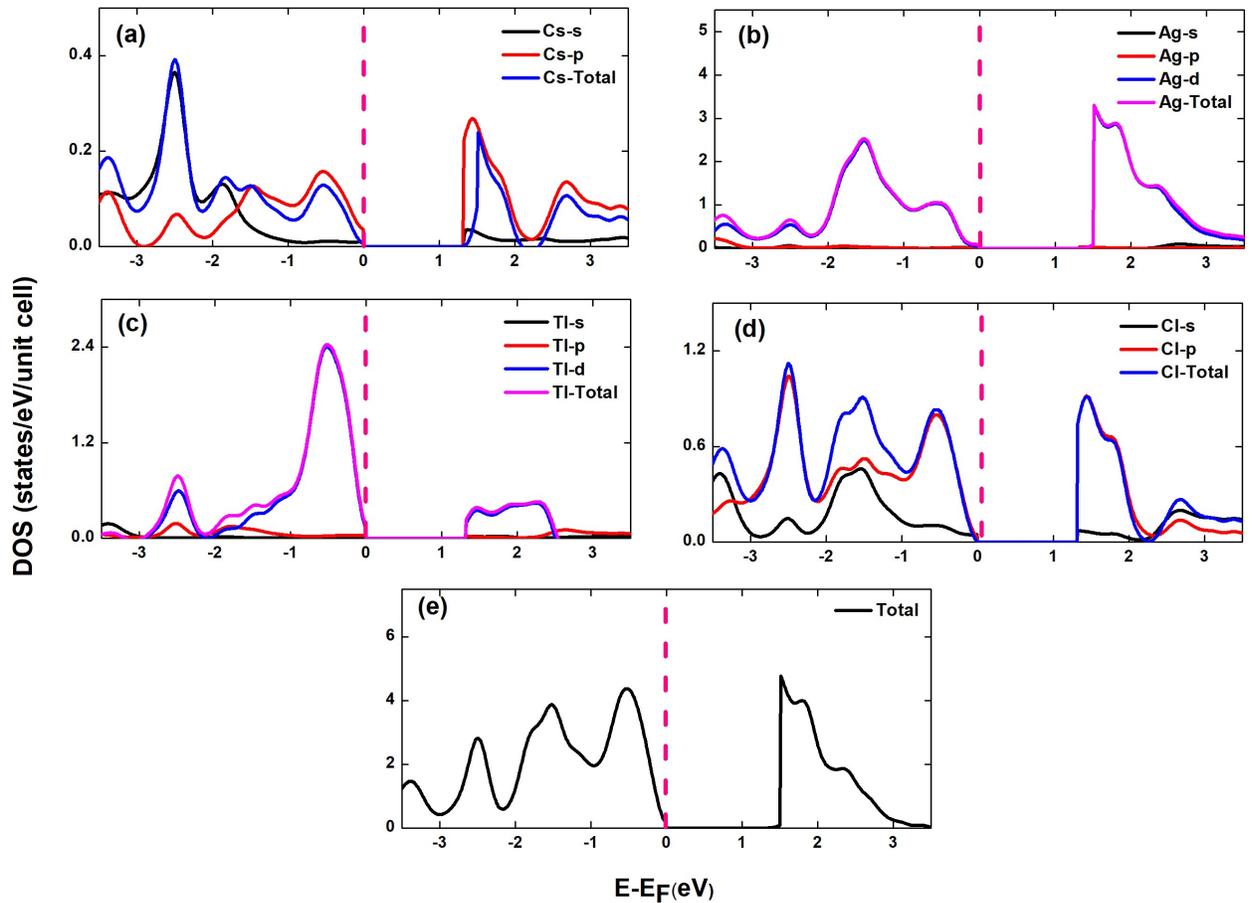

**Fig.3:** Total and projected DOS of $Cs_2AgTlCl_6$ using GGA+U method.

**Table 2:** Band-gaps comparison with reported theoretical/experimental data of elapsolite $Cs_2AgTlX_6$ (X= Cl, Br) by GGA+U approach.

| Compound | DFT Functional | Calculated Band Gap(eV) | Reference |
|---|---|---|---|
| $Cs_2AgTlCl_6$ | PBE+U | 1.32 | 1.96 (experimental)[18] |
| | | | 1.87 (theoretical) [18] |
| $Cs_2AgTlBr_6$ | PBE+U | 0.61 | 0.95 (experimental)[18] |
| | | | 0.63 (theoretical)[18] |

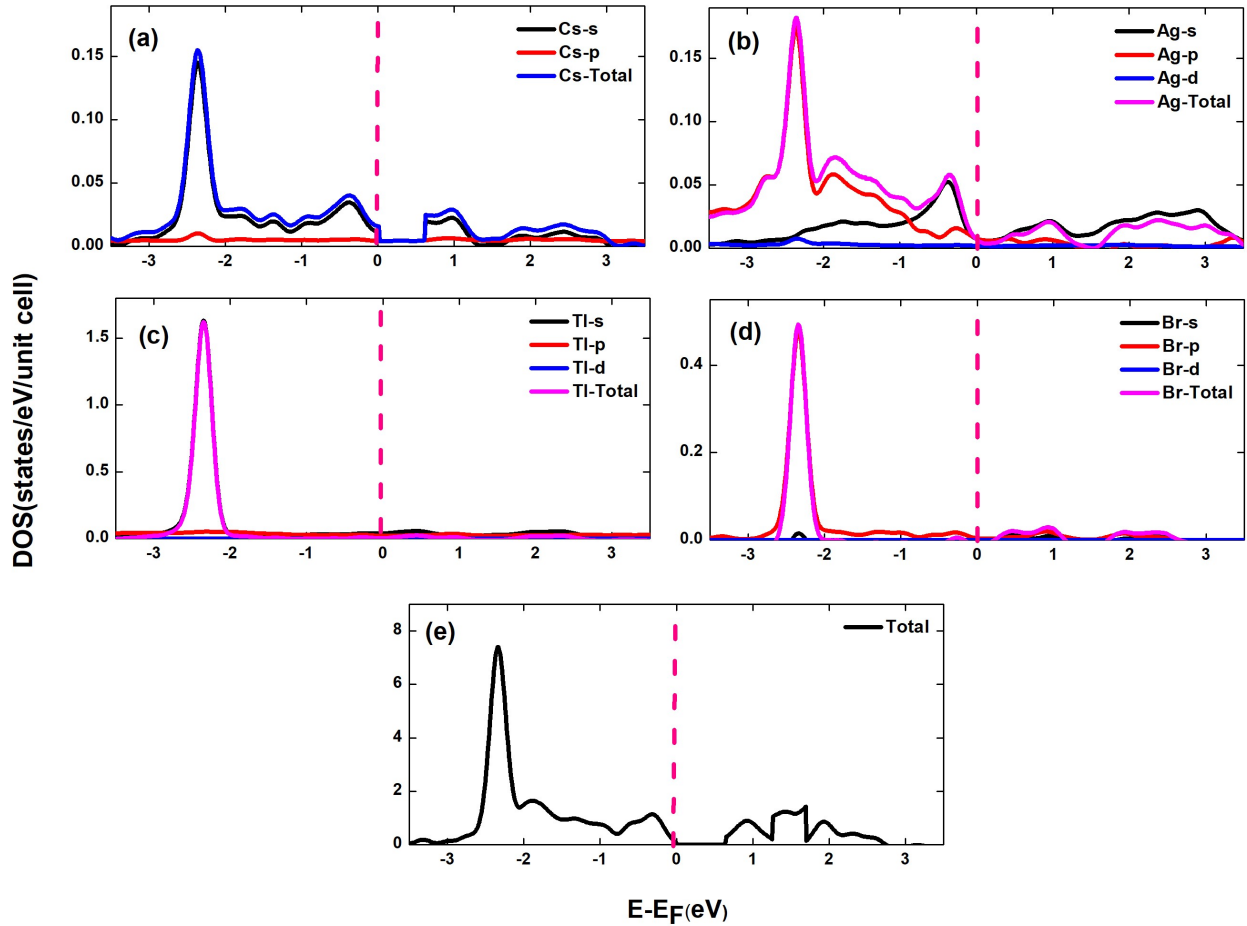

**Fig.4:** Total and projected DOS of $Cs_2AgTlBr_6$ using the GGA+U method.

The underlying physics behind the small band gaps can be interpreted based on the presence of diffused spherical orbitals (Tl-s and Ag-s) at conduction band minimum which in turn exhibits the halide orbital overlap leading to highly dispersive bands. These dispersive bands lead to asmall effective mass ($m_e$ =0.23)which causes significant mobility of electrons [18]. The valence band maxima and conduction band minima come very close to each other due to this highly dispersive nature of bands thus creating small band gaps.Also, a small gap is generated due to thesmall energy difference between filled Ag-d and Tl-s orbitals.

## 5. Optical properties

Photovoltaic device efficiency is a primary criterion in optoelectronics to determine the worth of photovoltaic devices for applications in solar cell fabrication for energy harvesting. In this scenario, thedielectric function $\epsilon(\omega)$ given by Ehren-reich and Cohen's relation [35] in equation (3), is an important complex function that determines how a photovoltaic material gives a response when exposed to anexternal electromagnetic field. The other optical parameters of interest such as absorption coefficient $\alpha(\omega)$, refractive index $n(\omega)$, reflectivity $R(\omega)$, extinction coefficient $K(\omega)$, optical conductivity $\sigma(\omega)$ and energy loss function $L(\omega)$ can be extracted using realand imaginary parts of dielectric function i.e. $\epsilon_1(\omega)$ and $\epsilon_2(\omega)$ using equations (4-9) [35-37].

$$\epsilon(\omega) = \epsilon_1(\omega) + i\,\epsilon_2(\omega) \qquad (3)$$

$$n(\omega) = \frac{\left[\{\epsilon_1^2(\omega) + \epsilon_2^2(\omega)\}^{1/2} + \epsilon_1(\omega)\right]^{1/2}}{\sqrt{2}} \qquad (4)$$

$$K(\omega) = \frac{\left[\{\epsilon_1^2(\omega) + \epsilon_2^2(\omega)\}^{1/2} - \epsilon_1(\omega)\right]^{1/2}}{\sqrt{2}} \qquad (5)$$

$$R(\omega) = \frac{(n-1)^2 + K^2}{(n+1)^2 + K^2} \qquad (6)$$

$$L(\omega) = \frac{\epsilon_2(\omega)}{\epsilon_1^2(\omega) + \epsilon_2^2(\omega)} \qquad (7)$$

$$\alpha(\omega) = \frac{2K\omega}{c} \qquad (8)$$

$$\sigma(\omega) = \frac{\alpha(\omega)n(\omega)c}{4\pi} \qquad (9)$$

The calculated values of theabove-mentioned optical parameters are plotted in Fig. 5 within the energy range of 0-20eV. The real part of thedielectric function $\epsilon_1(\omega)$ reveals the information about dispersion while the imaginary part $\epsilon_2(\omega)$ gives information about absorption. Both $\epsilon_1(\omega)$ and $\epsilon_2(\omega)$ have nearly identical behavior for both $Cs_2AgTlCl_6$ and $Cs_2AgTlBr_6$ as shown in Fig. 5(a) and 5(b) due to their similar electronic band structures. The $\epsilon_1(\omega)$ tells about the amount of light scattered and also the phase velocity that in turn depends on the maximum dispersion of light. The values of $\epsilon_1(\omega)$ when $\omega \to 0$ are ~5.61 and ~6.30 for $Cs_2AgTlCl_6$ and $Cs_2AgTlBr_6$ respectively as shown in Fig. 5(a). Penn's model is satisfied here i.e. the inverse relation between static dielectric constant $\epsilon_1(0)$ and band gap $E_g$ is satisfied [38-42].

$$\epsilon(0) \approx \epsilon_1(0) \approx 1 + \left(\frac{\hbar\omega}{E_g}\right)^2 \qquad (10)$$

The $\epsilon_2(\omega)$ showing the electronic transitions from valance to conduction band have prominent peaks at 0.06, 2.78, and 5.06eV for $Cs_2AgTlCl_6$ and 0.49, 0.93 and 1.75eV for $Cs_2AgTlBr_6$ as shown in Fig. 5(b).

The amount of light absorbed by a material is identified by its absorption coefficient α(ω). It is one of the most typical parameters for categorizing a material whether it is suitable or not for solar cell applications. It is evident from Fig. 5(c) that prominent optical transitions occur at ~2.41eV (514nm) and ~2.34eV (530nm) for $Cs_2AgTlCl_6$ and $Cs_2AgTlBr_6$ respectively. These transitions correspond to the green light of visible spectra. Both compounds have large values of α(ω) for a wide range of energies that also includes the visible range (~1.7eV-3eV) hence it can be assumed that these materials will be attractive for solar cell applications [43, 44].

Refractive index n(ω) is also an important parameter which determines that whether the material is transparent or opaque while investigating the optical properties. It is interesting to note that n(ω) follows a similar trend as that of $\epsilon_1(\omega)$ as shown in Fig. 5(d). The values of static refractive index $n_1(0)$ are 2.37 and 2.51 when $\omega \to 0$ for $Cs_2AgTlCl_6$ and $Cs_2AgTlBr_6$ respectively. The values of $n_1(0)$ obeys the relation $n_1(0)^2 \approx \epsilon_1(0)$. The other dominant peaks for n(ω) occur at 3.95, 5.26, and 12.18eV for $Cs_2AgTlCl_6$ and 4.87, 6.30, and 12.43eV for $Cs_2AgTlBr_6$. The reflectivity percentage is represented in Fig. 5(e) which is also one of the substantial parameters signifying the percentage of incident light reflected at a particular frequency. Initially, its values were high i.e.76.6% and 71.3% at $\omega = 0$ but it decreased with frequency, which is a better sign for these materials to be used in designing the optical devices. The minimum reflectivity in the visible range was 60.3% and 54.1% for $Cs_2AgTlCl_6$ and $Cs_2AgTlBr_6$ respectively. The extinction coefficient $K(\omega)$, also known as the imaginary refractive index, is closely related with $\epsilon_2(\omega)$ thus having a similar trend as illustrated in Fig. 5(f).

The highest values of $K(\omega)$ were observed at 9.82 and 9.51 for $Cs_2AgTlCl_6$ and $Cs_2AgTlBr_6$ respectively. The plot of optical conductivity $\sigma(\omega)$ versus energy is presented in Fig. 5(g). The highest values were observed as $6.88 \times 10^{16}$ sec$^{-1}$ and $4.09 \times 10^{16}$ sec$^{-1}$ at 0.20eV for $Cs_2AgTlCl_6$ and $Cs_2AgTlBr_6$ respectively. Moreover, reasonably good values ($\sim 10^{16}$ sec$^{-1}$) were observed in the visible range (~1.7eV-3eV). The photon energy lost while passing through the material was measured by energy loss function $L(\omega)$ as plotted in Fig. 5(h). The highest peaks were observed in the range of 6-8eV and 19-20eV for both compounds.

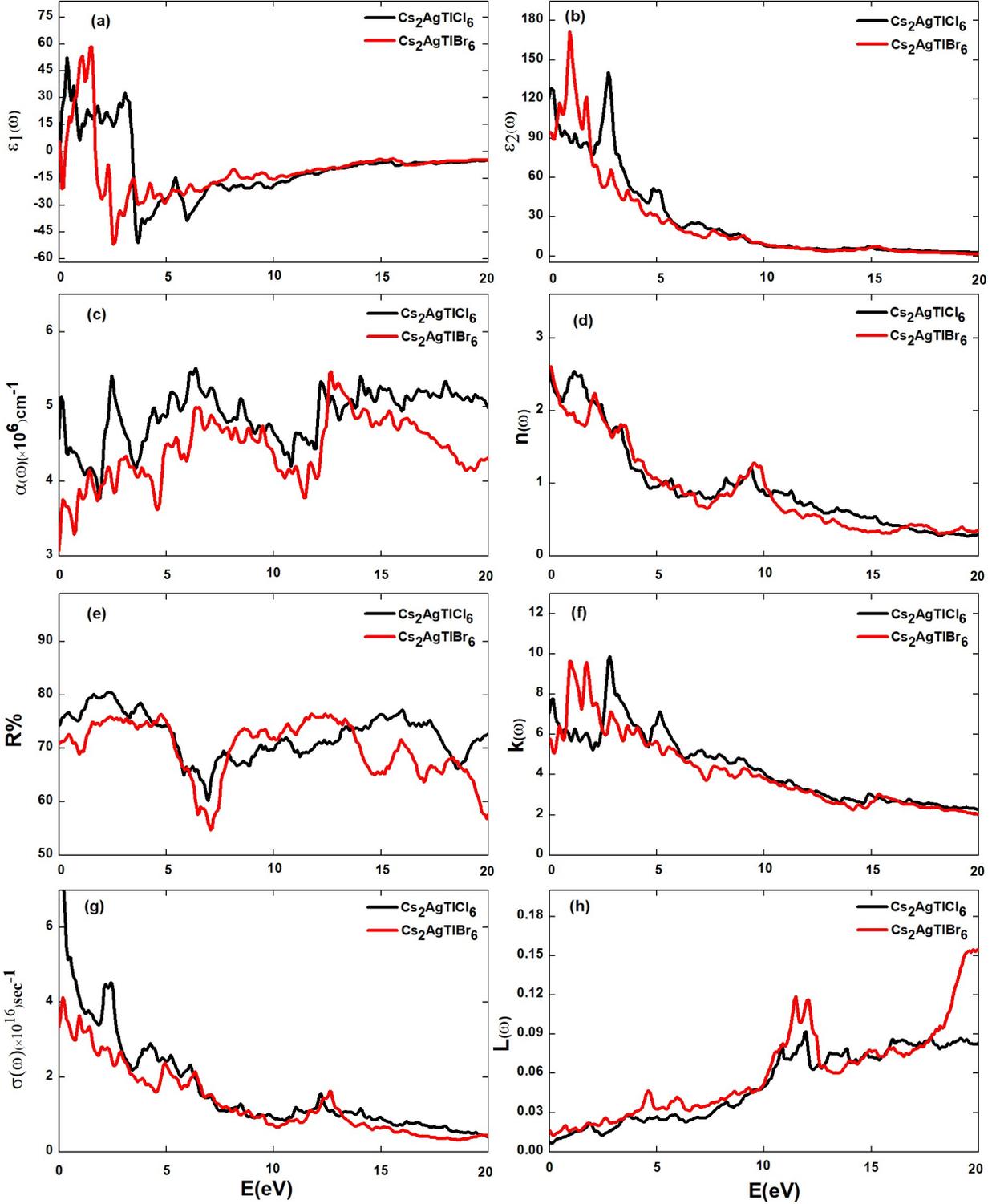

**Fig.5:** Variation of various optical parameters with energy in the range of 0-20eV.

## 6. Thermoelectric properties:

BoltzTrap was used to calculate thermoelectric properties, which is attributed to the Boltzmann transport theory for electronic transport within solids [24]. Currently, we focused on intrinsic carrier concentration within the framework of CRTA [45]. Thermoelectric parameters like electrical conductivity, electronic thermal conductivity, Seebeck coefficient, and power factorare introducedin Fig. 6with respect to thetemperature, whereas,specific heat capacity at constant volume, carrier charge density, Hall coefficient, and magnetic susceptibility are represented in Fig. 7 and figure of merit with respect to temperature are shown in Fig. 8 for both $Cs_2AgTlCl_6$ and $Cs_2AgTlBr_6$.Various parameters concerned with thermoelectric performance are not exclusive but rather depend on each other as well as on the electronic and structural properties of the compounds. The higher figure of merit which is the indication of a good thermoelectric material merely depends on electrical conductivity, Seebeck coefficient, and thermal conductivity. A higher electrical conductivity and lower thermal conductivity are desirable for the higher figure of merit. The power factor is the product of the Seebeck coefficient and electrical conductivity whereas both the Seebeck coefficient and electrical conductivity also depend on carrier concentration. Meanwhile, a lower value of thermal conductivity is required since it is inversely proportional to the figure of merit. It is observed that a small band gap material with dispersive bands has low effective mass, high electrical conductivity, and a large Seebeck coefficient. Therefore, the materials under study are a good choice as thermoelectric materials.

Electrical conductivity per unit relaxation time ($\sigma/\tau$) versus temperature is plottedin the 0-800K temperature range as displayed in Fig. 6(a). It incorporates the response of charge carriers upon a temperature gradient and it is noticeable in our case that $\sigma/\tau$ shows a gradual increase having the highest values of $3.02 \times 10^{21} (\Omega ms)^{-1}$ and $3.58 \times 10^{21} (\Omega ms)^{-1}$ at 800K for $Cs_2AgTlCl_6$ and $Cs_2AgTlBr_6$respectively. The increasing trend reveals the semiconducting nature of both elapsolite. This is because of theprogress of kinetic energy of charge carriers with apositive temperature gradient in a semiconductor and this property plays a vital role in thethermoelectric performance of a compound [46].

The electronic contribution of thermal conductivity per unit relaxation time ($\kappa_e/\tau$) is displayed in Fig. 6 (b). It has amonotonic increasing trend for both of the compounds with themaximum value of $5.90\times10^{11}$ W/mKs and $1.38\times10^{11}$ W/mKs for $Cs_2AgTlCl_6$ and $Cs_2AgTlBr_6$ at 800Krespectively. In accordance with Wiedemann-Franz law i.e. ratio of thermal to electrical conductivity remains constant ($\kappa_e/\sigma$ =LT) is obeyed here [47]. The obtained values are in anintermediate range whicha good trend is since higher valuesof thermal conductivitiesdecrease the efficiency of thermoelectric materials.

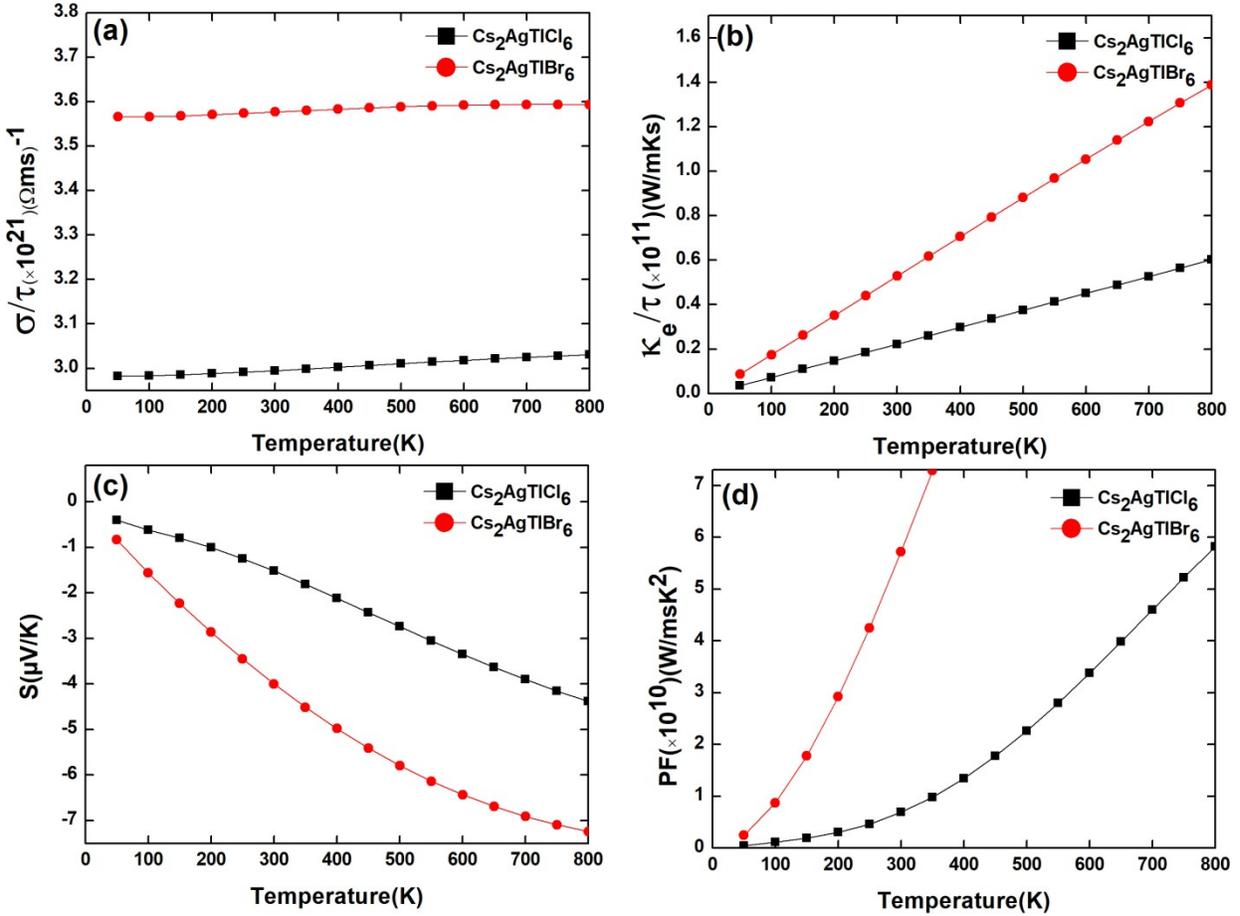

**Fig.6:** Temperature-dependent (a) electrical conductivity ($\sigma/\tau$), (b) electronic thermal conductivity ($\kappa_e/\tau$), (c) Seebeck coefficient (S), and (d) power factor (PF).

The ratio of potential difference created tothe applied temperature gradient defines the Seebeck coefficient (S) which is one of the most prominent parameters in thermoelectric investigations. In thepresent study, the negative values of S convey the N-type semiconducting behavior of $Cs_2AgTlCl_6$ and $Cs_2AgTlBr_6$ [46]. The decreasing trend is observed with maximum values of -0.41μV/K and -0.82μV/K for $Cs_2AgTlCl_6$ and $Cs_2AgTlBr_6$respectively as shown in Fig. 6(c).The performance of thermoelectric material is assessed by its power factor defined as PF=$S^2\sigma/\tau$. For these compounds, it shows a rapidly increasing behavior as illustrated in Fig. 6(d) with thehighest value of $5.74\times10^{10}$ W/msK$^2$ at 800K for Cs$_2$AgTlCl$_6$ while $7.20\times10^{10}$ W/msK$^2$ at 350K for Cs$_2$AgTlBr$_6$.

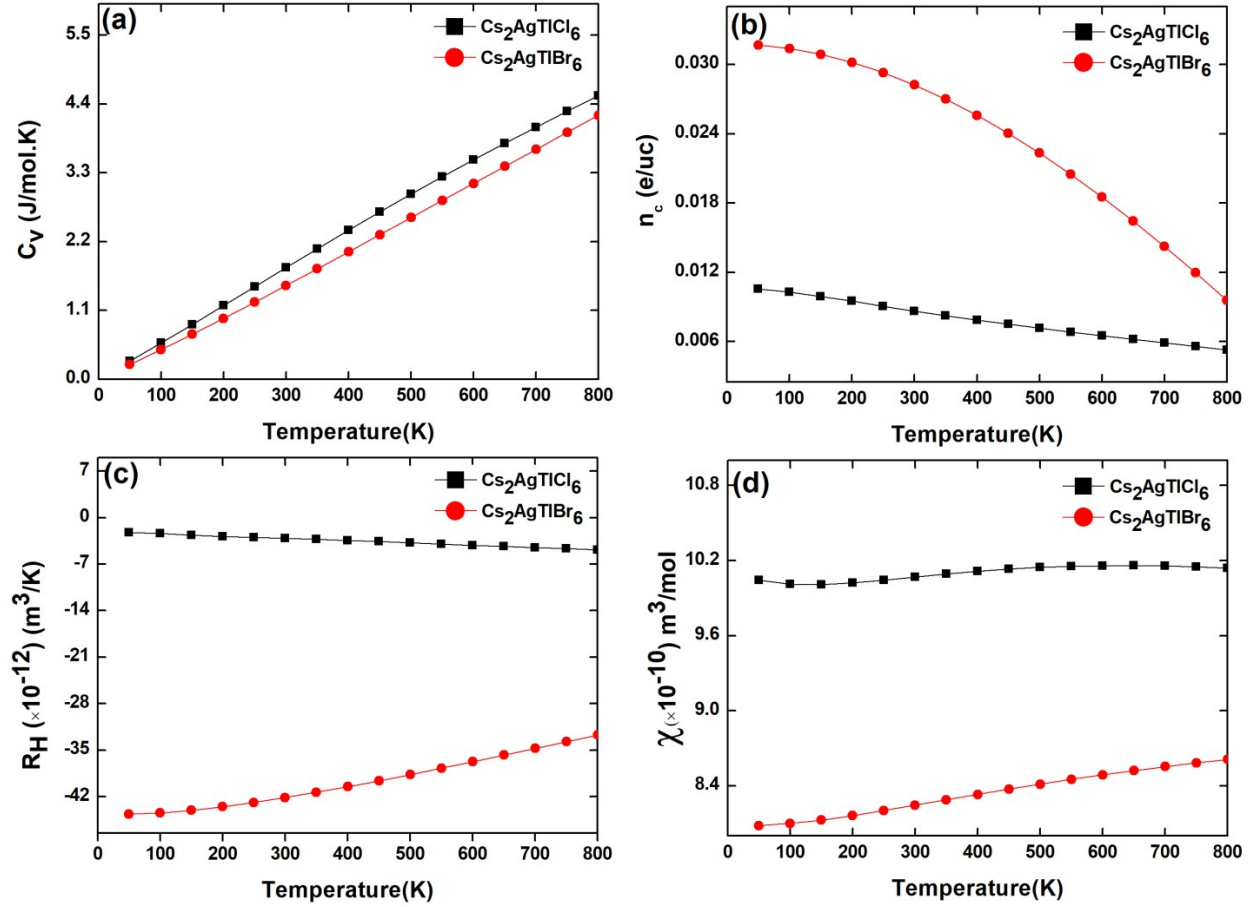

**Fig.7:** Temperature-dependent (a) specific heat capacity at constant volume (C$_v$), (b) carrier charge density (n$_c$), (c) Hall coefficient (R$_H$), and (d) magnetic susceptibility ($\chi$).

The heat absorption capability of a material can be evaluatedusing specific heat which is the heat required to make a unit temperature increment for theunit mole of a compound. In our case, specific heat capacity at constant volume (C$_v$) is increasing with the temperature which is in accordance with theDebye model of specific heat (C$_v$ $\alpha$ T$^3$). The highest observed value was 4.47 J/mol.K and 4.17 J/mol.K at 800K for Cs$_2$AgTlCl$_6$ and Cs$_2$AgTlBr$_6$respectively as depicted in Fig. 7(a).The variation in carrier charge density (n$_c$) with respect to the temperature is plotted in Fig. 7(b) showing a gradual decrease in n$_c$ for both compounds. The maximum values were 0.01 e/uc and 0.03 e/uc at 50K for Cs$_2$AgTlCl$_6$ and Cs$_2$AgTlBr$_6$respectively. It incorporates the transfer of electrons from valance to conduction bands. Hall coefficient ($R_H$) is used to collect information about the type and number of charge carriers contributing to total electric current. In Fig. 7(c), $R_H$ versus T graph shows that it remains almost constant ~ $-2.2\times10^{-12}$ m$^3$/K and ~ $-3.2\times10^{-11}$ m$^3$/K for $Cs_2AgTlCl_6$ and $Cs_2AgTlBr_6$. The negative values depict that the majority of charge carriers are electrons further confirming that both of the compounds are N-type semiconductors.

The graph of magnetic susceptibility ($\chi$) with temperature also remains almost constant as shown in Fig. 7(d). The values are ~$1.0\times10^{-9}$ m$^3$/mol and ~$8.41\times10^{-10}$ m$^3$/mol for $Cs_2AgTlCl_6$ and $Cs_2AgTlBr_6$, respectively.

The dimensionless figure of merit defined below in equation (11) predicts the proficiency of a material for its usage in various thermoelectric applications.

$$ZT = \frac{S^2 \sigma T}{\kappa_{tot}} \quad (11)$$

Here the total thermal conductivity $\kappa_{tot}$ has a contribution from electronic and phononic transport i.e. $\kappa_{tot} = \kappa_e + \kappa_{ph}$. Lattice thermal conductivity ($\kappa_{ph}$) has an inverse relation with temperature ($\kappa_{ph} \propto T^{-1}$)[47,48], hence it can be concluded that it plays a less significant role at higher temperatures in semiconductors. This is the reason; we used the BoltzTrap code that calculates only electronic contributions. Moreover, both electrical and thermal conductivity relies on relaxation time $\tau$ but as ZT depends on the ratio of electrical to thermal conductivity hence relaxation time is canceled that's why if we consider it to be constant, it doesn't affect the value of ZT. The value of $\tau$ is usually in the range of $10^{-13}$-$10^{-14}$ seconds [49] and here we considered $\tau = 1\times10^{-14}$ seconds based on constant relaxation time approximation [25].

In our case, we have plotted the electronic part of the figure of merit ($ZT_e$) versus temperature in the range of 0-800K as shown in Fig. 8. The maximum values of $ZT_e$ were observed as 0.96 at 800K and 1.41 at 600K for $Cs_2AgTlCl_6$ and $Cs_2AgTlBr_6$ respectively. These obtained values of $ZT_e$ closer to or greater than unity categorize the compounds under investigation in the class of efficient thermoelectric materials [25], being very attractive for thermoelectric generators and other relevant applications.

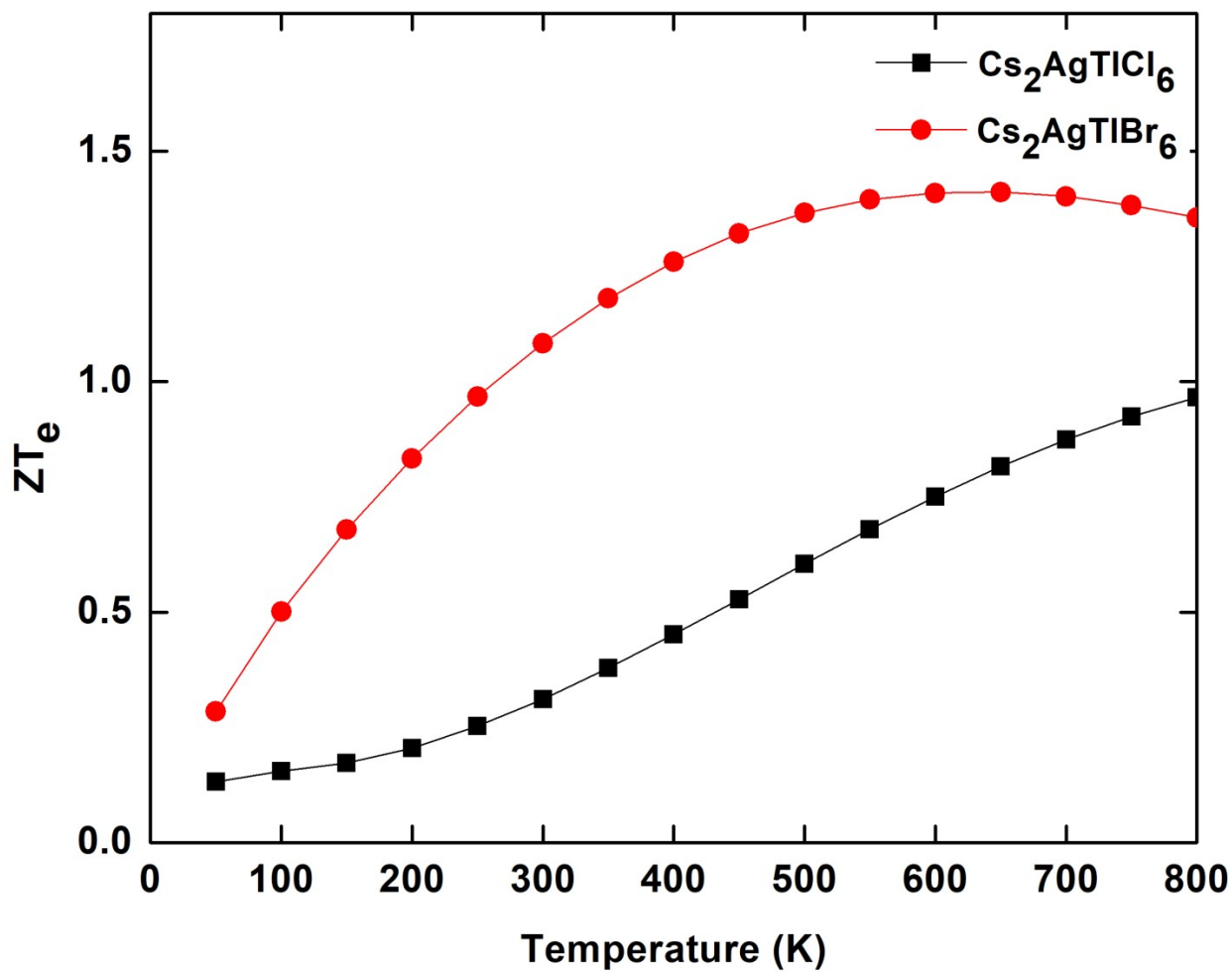

**Fig.8:** Representation of the electronic part of the figure of merit ($ZT_e$) with respect to the temperature.

## 7. Conclusion:

In brief concluding remarks, a comprehensive DFT study of the structural, electronic, phononic, optical, and thermoelectric properties of elapsolite $Cs_2AgTlX_6$ (X= Cl, Br) was conducted. The Perdew, Burke, and Ernzerhof (PBE) exchange-correlation potential with the Hubbard-U correction was used. Herein, both $Cs_2AgTlCl_6$ and $Cs_2AgTlBr_6$ are found to be stable double perovskites with estimated values of Goldschmidt's factors in thestable structural range i.e. the value of $\tau_G$ are 0.85 and 0.81 while that of μ are 0.69 and 0.63. $Cs_2AgTlX_6$ (X= Cl, Br) are direct band gap semiconductors with values of 1.32eV and 0.61eV. Prominent absorption peaks and significant optical conductivity (~$10^{16}$ sec$^{-1}$) in thevisible range (~1.7eV-3eV)were observed suggesting various optoelectronic applications.

Despite this, the negative values of the Seebeck coefficient indicate the N-type semiconducting behavior of both compounds. The Power factors of the compounds under study were significant i.e. $5.74\times10^{10}$ W/msK$^2$ for $Cs_2AgTlCl_6$ while $7.20\times10^{10}$ W/msK$^2$ for $Cs_2AgTlBr_6$. The maximum values of $ZT_e$ are 0.96 at 800K and 1.41 at 600K for $Cs_2AgTlCl_6$ and $Cs_2AgTlBr_6$ respectively. The small direct band gaps, structural and thermodynamic stability, high absorption coefficients and optical conductivities, large Seebeck coefficients, significant power factors, and figure of merits close to unity suggest that both of the compounds are highly appropriate for various optoelectronic and thermoelectric device applications.

**Acknowledgment:**

The authors acknowledge the computing facilities provided by the Centre for High-Performance Computing (CHPC-MATS1424), Cape Town, South Africa.

[7].    Hussain M, Rashid M, Ali A, Bhopal MF, Bhatti AS. Systematic study of optoelectronic and transport properties of cesium lead halide ($Cs_2PbX_6$; X= Cl, Br, I) double perovskites for solar cell applications. *Cer Int.* 2020;46(13):21378-87.

[8].    Zhao XG, Yang JH, Fu Y, Yang D, Xu Q, Yu L, Wei SH, Zhang L. Design of lead-free inorganic halide perovskites for solar cells via cation-transmutation. *J of the American Chem Soc.* 2017;139(7):2630-8.

[9].    Jung EH, Jeon NJ, Park EY, Moon CS, Shin TJ, Yang TY, Noh JH, Seo J. Efficient, stable and scalable perovskite solar cells using poly (3-hexylthiophene). *Nature.* 2019;567(7749):511-5.

[10].   NREL N. Best research-cell efficiencies. National Renewable Energy Laboratory: Golden, Colorado. 2019.

[11].   Nicholson KM, Kang SG, Sholl DS. First principles methods for elpasolite halide crystal structure prediction at finite temperatures. *J of All and Comp.* 2013;577:463-8.

[12].   Chu L, Ahmad W, Liu W, Yang J, Zhang R, Sun Y, Yang J, Li XA. Lead-free halide double perovskite materials: a new superstar toward green and stable optoelectronic applications. *Nano-Micro Lett.* 2019;11(1):1-8.

[13].   Filip MR, Hillman S, Haghighirad AA, Snaith HJ, Giustino F. Band gaps of the lead-free halide double perovskites $Cs_2BiAgCl_6$ and $Cs_2BiAgBr_6$ from theory and experiment. *The J of phychem Lett.* 2016;7(13):2579-85.

[14].   Slavney AH, Leppert L, Bartesaghi D, Gold-Parker A, Toney MF, Savenije TJ, Neaton JB, Karunadasa HI. Defect-induced band-edge reconstruction of a bismuth-halide double perovskite for visible-light absorption. *J of the American Chem Society.* 2017;139(14):5015-8.

[15].   Wei F, Deng Z, Sun S, Hartono NT, Seng HL, Buonassisi T, Bristowe PD, Cheetham AK. Enhanced visible light absorption for lead-free double perovskite $Cs_2AgSbBr_6$. *Chem Communications*. 2019;55(26):3721-4.

[16].   Volonakis G, Haghighirad AA, Milot RL, Sio WH, Filip MR, Wenger B, Johnston MB, Herz LM, Snaith HJ, Giustino F. $Cs_2InAgCl_6$: a new lead-free halide double perovskite with direct band gap. The J of PhyChem Lett. 2017;8(4):772-8.

[17].   Prasanna R, Gold-Parker A, Leijtens T, Conings B, Babayigit A, Boyen HG, Toney MF, McGehee MD. Band gap tuning via lattice contraction and octahedral tilting in perovskite materials for photovoltaics. *J of the American Chem Society*. 2017;139(32):11117-24.
20[7].    Hussain M, Rashid M, Ali A, Bhopal MF, Bhatti AS. Systematic study of optoelectronic and transport properties of cesium lead halide ($Cs_2PbX_6$; X= Cl, Br, I) double perovskites for solar cell applications. *Cer Int.* 2020;46(13):21378-87.

[8].    Zhao XG, Yang JH, Fu Y, Yang D, Xu Q, Yu L, Wei SH, Zhang L. Design of lead-free inorganic halide perovskites for solar cells via cation-transmutation. *J of the American Chem Soc.* 2017;139(7):2630-8.

[9].    Jung EH, Jeon NJ, Park EY, Moon CS, Shin TJ, Yang TY, Noh JH, Seo J. Efficient, stable and scalable perovskite solar cells using poly (3-hexylthiophene). *Nature.* 2019;567(7749):511-5.

[10].   NREL N. Best research-cell efficiencies. National Renewable Energy Laboratory: Golden, Colorado. 2019.

[11].   Nicholson KM, Kang SG, Sholl DS. First principles methods for elpasolite halide crystal structure prediction at finite temperatures. *J of All and Comp.* 2013;577:463-8.

[12].   Chu L, Ahmad W, Liu W, Yang J, Zhang R, Sun Y, Yang J, Li XA. Lead-free halide double perovskite materials: a new superstar toward green and stable optoelectronic applications. *Nano-Micro Lett.* 2019;11(1):1-8.

[13].   Filip MR, Hillman S, Haghighirad AA, Snaith HJ, Giustino F. Band gaps of the lead-free halide double perovskites $Cs_2BiAgCl_6$ and $Cs_2BiAgBr_6$ from theory and experiment. *The J of phychem Lett.* 2016;7(13):2579-85.

[14].   Slavney AH, Leppert L, Bartesaghi D, Gold-Parker A, Toney MF, Savenije TJ, Neaton JB, Karunadasa HI. Defect-induced band-edge reconstruction of a bismuth-halide double perovskite for visible-light absorption. *J of the American Chem Society.* 2017;139(14):5015-8.

[15].   Wei F, Deng Z, Sun S, Hartono NT, Seng HL, Buonassisi T, Bristowe PD, Cheetham AK. Enhanced visible light absorption for lead-free double perovskite $Cs_2AgSbBr_6$. *Chem Communications*. 2019;55(26):3721-4.

[16].   Volonakis G, Haghighirad AA, Milot RL, Sio WH, Filip MR, Wenger B, Johnston MB, Herz LM, Snaith HJ, Giustino F. $Cs_2InAgCl_6$: a new lead-free halide double perovskite with direct band gap. The J of PhyChem Lett. 2017;8(4):772-8.

[17].   Prasanna R, Gold-Parker A, Leijtens T, Conings B, Babayigit A, Boyen HG, Toney MF, McGehee MD. Band gap tuning via lattice contraction and octahedral tilting in perovskite materials for photovoltaics. *J of the American Chem Society*. 2017;139(32):11117-24.